\documentclass[useAMS,usenatbib]{mnras}
\usepackage{graphicx,psfig,cite,epsf}  
\usepackage{times}
\usepackage{subfigure}
\usepackage{textcomp}
\usepackage{multirow}
\usepackage{amsmath}	
\usepackage{amssymb}	
\usepackage{gensymb}

\def\apjl{ApJ}                
\def\apjs{ApJS}               
\def\ao{Appl.~Opt.}           
\def\aap{A\&A}                

\def\src{SAX J1748.9-2021}
\def\xmm{{\it XMM-Newton}}
\def\nustar{{\it NuSTAR}}
\def\inte{{\it INTEGRAL}}
\def\swift{{\it Swift/XRT}}

\title[A faint outburst of SAX J1748.9-2021]{A faint outburst of the accreting millisecond X-ray pulsar SAX J1748.9-2021 in NGC 6440}
\author[Pintore et al.] {Pintore F. $^{1}$\thanks{pintore@iasf-milano.inaf.it}, Sanna A.$^2$, Riggio A.$^2$, Di Salvo T.$^3$, Mereghetti S.$^1$, Bozzo E.$^4$, \newauthor S\'anchez-Fern\'andez C.$^5$, Burderi L.$^2$, Iaria R.$^3$\\
$^1$ INAF-Istituto di Astrofisica Spaziale e Fisica Cosmica - Milano, via E. Bassini 15, I-20133, Milano, Italy\\
$^2$ Universit\`a degli Studi di Cagliari, Dipartimento di Fisica, SP Monserrato-Sestu, KM 0.7, I-09042, Monserrato, Italy \\
$^3$  Dipartimento di Fisica e Chimica, Universit\`a di Palermo, via Archirafi 36, I-90123, Palermo, Italy \\
$^4$ ISDC, Department of Astronomy, University of Geneva, Chemin d\'Ecogia 16, 1290, Versoix, Switzerland\\ 
$^5$ European Space Astronomy Centre (ESA/ESAC), Science Operations Department, 28691, Villanueva de la Ca\~nada, Madrid, Spain}

\begin{document}

\maketitle

\begin{abstract}
\src\ is an accreting X-ray millisecond pulsar observed in outburst five times since its discovery in 1998. In early October 2017, the source started its sixth outburst, which lasted only $\sim13$ days, significantly shorter than the typical 30 days duration of the previous outbursts. It reached a 0.3--70 keV unabsorbed peak luminosity of $\sim3\times10^{36}$ erg s$^{-1}$. This is the weakest outburst ever reported for this source to date. We analyzed almost simultaneous \xmm, \nustar\ and \inte\ observations taken during the decaying phase of its 2017 outburst. We found that the spectral properties of \src\ are consistent with an absorbed Comptonization plus a blackbody component. The former, characterized by an electron temperature of $\sim20$ keV, a photon index of $\sim$1.6--1.7 keV and seed photon temperature of 0.44 keV, can be associated to a hot corona or the accretion column, while the latter is more likely originating from the neutron star surface (kT$_{bb}\sim0.6$ keV, R$_{bb}\sim2.5$ km). These findings suggest that \src\ was observed in a {\it hard} spectral state, as it is typically the case for accreting millisecond pulsars in outburst.
\end{abstract}


\section{Introduction}
\label{intro}

Accreting neutron stars (NS) with spin pulsations of the order of the millisecond ($\sim$1.6--10 ms) and hosted in low mass X-ray binary systems (LMXBs) belong to the class of the accreting X-ray millisecond pulsars (AMXPs), that counts 21 sources up to date \citep[e.g.][]{patruno17,sanna18,strohmayer18}. It has been proven that AMXPs are formed through the so-called {\it recycling} scenario (see e.g. \citealt{archibald09}, \citealt{papitto13b}b, for the observational evidences of such process). According to this scenario, the period of old pulsars in binary systems can be re-accelerated to a few milliseconds from the torque exerted by matter in the accretion disc formed during the outbursts. The properties of AMXPs suggest that NS magnetic fields of the order of $10^8-10^{9}$G \citep[e.g.][]{hartman08} are capable to drive a portion of the accreting matter from the accretion disc to the magnetic polar caps of the compact object, heating its surface. The photons produced by this process are responsible for the pulsating X-ray emission (assuming that the NS magnetic and rotational axes are misaligned, as it is often the case).
The timing analysis of the pulsations can provide information about the strength of the NS magnetic fields and the inner radius of the accretion discs (e.g. \citealt{burderi06, burderi07,burderi98,ghosh08,patruno09b,wilkinson11} or \citealt{disalvo08} for a review). The latter can be also inferred from either the detection of disc emission or, indirectly, from broad emission features (the most diffuse feature is the K-shell transition iron line at 6.4--7.0 keV) in the source X-ray spectrum. These are likely produced by reflection off of hard photons from the surface of the accretion disc and are affected by special and general relativistic effects caused by the fast rotation of the accretion disc close to the strong gravitational field of the compact object (e.g. \citealt{fabian89,cackett10,papitto09}; Di Salvo et al. 2018, submitted, and references therein). 

The spectral properties of AMXPs have been largely investigated (see e.g. \citealt{patruno12,burderi13,campana18}, and reference therein, for recent reviews) and showed that AMXPs are usually in {\it hard} states, characterized by the combination of a thermal soft component (temperatures $<2$ keV), a dominating Comptonizating component with electrons in an optically thin hot plasma with temperatures of 20--50 keV, and in some cases of a third thermal continuum component likely produced by the NS surface (e.g. \citealt{gilfanov98,gierlinski05,falanga05,patruno09b,papitto10}; \citealt{papitto13}a). It is believed that the first soft component is the accretion disc, while the hard component may be arising form either a hot corona or the accretion columns or the boundary layer \citep[e.g.][]{popham01}.

In this work, we focus on the AMXP SAX J1748.9-2021, discovered by {\it Beppo-SAX} in 1998 during its first recorded outburst. The source is located in the globular cluster NGC 6440 \citep{intZand99}, at a distance of $\sim8.5$ kpc and 0.6 kpc above the Galactic plane \citep{martins80,ortolani94,kuulkers03,valenti07}. \src\ experienced further outbursts in 2001, 2005, 2010 and 2015 \citep{intZand99,intZand01, verbunt00,markwardt05,patruno09, pintore16,sanna16}, with also a possible {\it a posteriori} associated outburst in 1971 \citep[][]{markert75}. During quiescence, the companion star was detected \citep{intZand01} and its mass was estimated to be in the range 0.1--1 M$_{\odot}$ \citep{altamirano08}. The binary system has a period of $\sim$8.76 hr and a projected semi-major axis of $\sim0.4$ light-seconds \citep{altamirano08, patruno09}.
\src\ showed intermittent pulsations at $\sim$442.361 Hz \citep{altamirano08, patruno09,sanna16} and it is known to emit numerous type-I X-ray bursts (observed with RXTE and \xmm; e.g. \citealt{galloway08,pintore16}). During the 2015 outburst, the source was observed in a soft state, with average spectral properties consistent with two soft thermal components plus a cold thermal Comptonized component ($\sim$2 keV) and an additional hard X-ray emission described by a power-law ($\Gamma\sim2.3$, \citealt{pintore16}). 
These components were associated to the accretion disc, the NS surface and a thermal Comptonized emission coming out of an optically thick plasma region, respectively. The origin of the high energy tail was unclear, although a similar component has been detected in several other LMXBs during soft states \citep[e.g.][]{disalvo00,paizis06,dai07,tarana07,piraino07}.

In 2017, \src\ underwent its sixth outburst. The event was firstly detected by MAXI/GSC \citep{negoro17} on September, 29th and then observed by \swift, \inte, \xmm\ and \nustar\ \citep{bahramian17,harita17,digesu17}. Here we present the study of the 2017 broad-band spectral properties of \src\ using \xmm, \nustar, \inte\ and \swift\ observations.

\section{Data Reduction}
\label{data_reduction}

\begin{figure}
\center
\includegraphics[width=8.7cm]{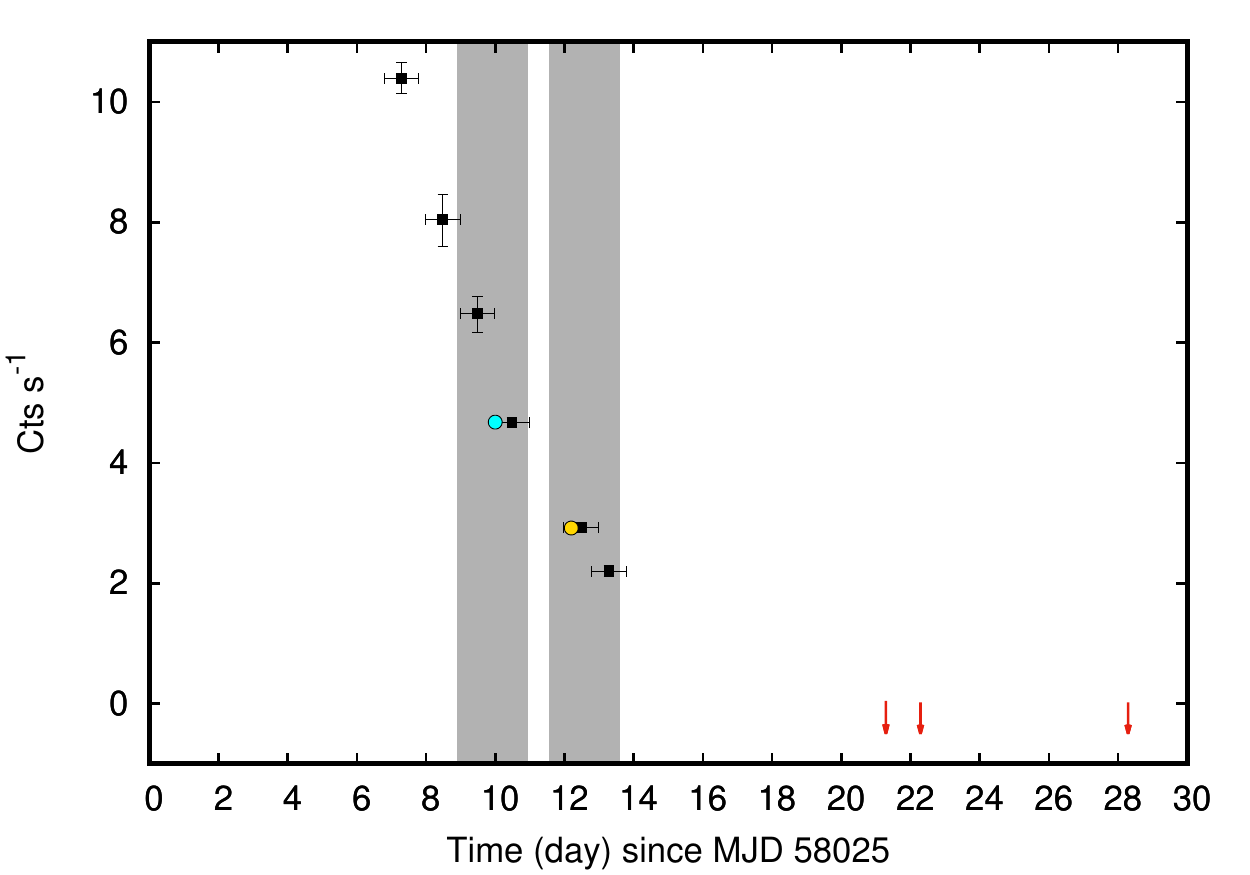}
\includegraphics[width=8.5cm]{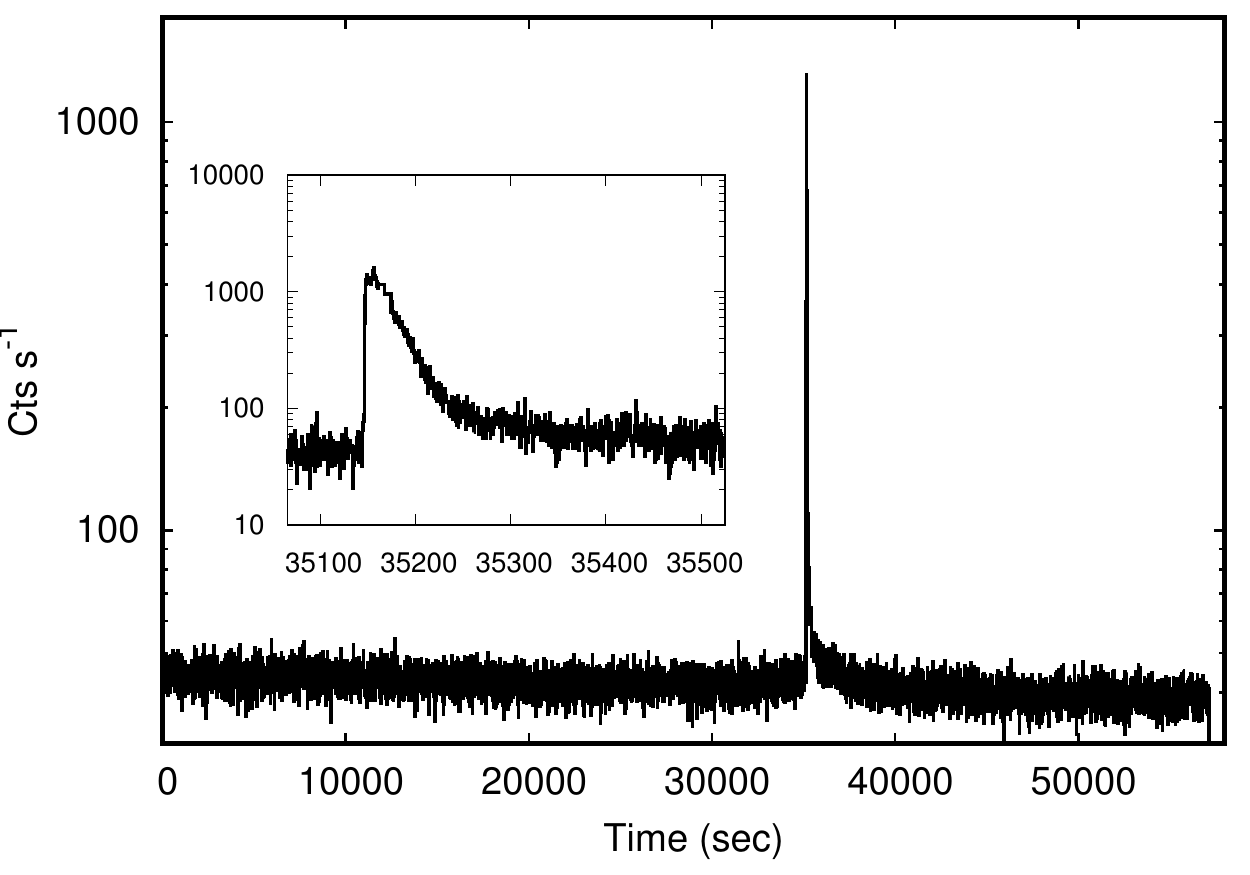}
\caption{Top: background subtracted 0.3--10 keV {\it Swift-XRT} lightcurve (starting from the beginning of the outburst as reported by MAXI), where red arrows are $3\sigma$ upper limits; the \xmm\ and \nustar\ observations have been taken at MJD 58035 (cyan circle) and 58037 (orange circle), respectively. \inte\ observations were taken in the epochs inside the grey boxes. Bottom: background subtracted 0.3--10 keV {\it XMM/EPIC-pn} lightcurve; a type-I burst (zoomed in the inset) happened at $\sim35$ ks after the beginning of the observation.}
\label{lc_xmm}
\end{figure}

We analysed \xmm\ and \nustar\ DDT observations performed on October, 9th 2017 (Obs.ID. 0795712201) and October, 11th 2017 (Obs.ID. 90301320002), respectively. We also investigated all the \swift\ observations which monitored the source outburst (Figure~\ref{lc_xmm}-top), as well as all the available \inte\ observation taken between October, 7th and 13th. 
\newline
\newline
\noindent {\bf XMM-Newton} The \xmm\ observation was taken in TIMING mode for a total exposure of $\sim$56 ks. We extracted EPIC-pn \citep{struder01short} events only as the two EPIC-MOS cameras \citep{turner01short} were not pointing the source. We reduced the EPIC-pn data with SAS v15.0.0 (using the RDPHA corrections, e.g. \citealt{pintore14b}), selecting single- and double-pixel events ({\sc pattern}$\leq$4). We extracted source and background events from RAWX=[32:44] and RAWX=[3:5], respectively. We verified that the background extracted in this region was not heavily contaminated by the source, comparing it with the background estimated from the EPIC-MOS data. During the observation, \src\ showed a type-I burst (Figure~\ref{lc_xmm}-bottom).

As the average EPIC-pn count rate was $\sim30$ cts s$^{-1}$, it was well below the threshold for pile-up\footnote{https://heasarc.gsfc.nasa.gov/docs/xmm/sas/USG/epicpileuptiming.html}. We extracted the average EPIC-pn spectrum and rebinned it with an oversample of 3 channels per energy resolution element using the \textit{specgroup} task. The spectrum was analyzed in the range 1.3--10 keV, to limit the EPIC-pn calibration uncertainties at low energies when operated in TIMING mode.

The RGS spectra, first and second order, were extracted using the standard \textit{rgsproc} task. We created an average spectrum merging RGS1 and RGS2 data (using the tool \textit{rgscombine} which combines, separately, the first and second order spectra). We rebinned the final spectrum with at least 100 counts per bin and we fitted it in the range 0.6--2.0 keV
\newline
\newline
\noindent {\bf NuSTAR} We followed the standard procedure based on {\sc NUSTARDAS} (the \nustar\ {\it  Data Analysis Software} v1.3.0) in the HEASOFT {\sc ftools} v6.23 to reduce the \nustar\ data and obtain cleaned event files. The total exposure time was 44 ks.
We extracted source and background events from circular regions of radii 60'' and 90'', respectively. No type-I bursts were revealed during the observation. The FMPA and FMPB spectra were then grouped to have at least 100 counts per bin, respectively, and we fitted them in the range 3--70 keV.
\newline
\newline
\noindent {\bf Swift} We used all the available 9 \swift\ observations collected between 2017, October 5 and 14, in both PC (the first and the last three observations of the monitoring) and WT mode (5 observations between 7--11 October). We followed standard procedures to reduce the data and extract spectra and lightcurves\footnote{http://www.swift.ac.uk/analysis/xrt/index.php}. We grouped the spectra to have at least 25 counts per bin and we fitted them in the range 0.3--10 keV. During the first {\it Swift} observation (taken in PC mode), pile-up was not negligible and we corrected for its effect \footnote{following the thread in http://www.swift.ac.uk/analysis/xrt/pileup.php} when we analyzed this data.
\newline
\newline
\noindent {\bf INTEGRAL} We also analyzed the available \inte\ observations of the field around \src\ during the satellite revolutions 1871 and 1872. These covered the time span from 2017 October 7 at 21:39 to October 9 at 22:47 (UTC) and from 2017 October 10 at 13:29 to October 12 at 14:49 (UTC), respectively. We analyzed all the publicly available \inte\ data and the data for which our group got data rights in AO14 by using version 10.2 of the Off-line Scientific Analysis software (OSA) distributed by the ISDC \citep{courvoisier03}. \inte\ observations are divided into ``science windows'' (SCWs), i.e. pointings with typical durations of $\sim$2-3 ks. 

The source was within the JEM-X field of view \citep[FoV; ][]{lund03} only for 6 SCWs during revolution 1872 and the total effective exposure time was of only 12.9 ks. As the source was located at the rim of the instrument FoV, the statistics of the data was too low to extract a meaningful spectrum. We used the JEM-X1 and JEM-X2 lightcurves with a time resolution of 2~s to search for type-I X-ray bursts, but no significant detections were found.  

We used for IBIS/ISGRI \citep{ubertini03,lebrun03} all SCWs where the source was located within 12 degree from the center of the FoV. This provided a total exposure of 58.4 ks and 44.6 ks in revolution 1871 and 1872, respectively. 
We extracted two ISGRI spectra, one for each revolution. Both these spectra could be well fit with a simple power-law with photon index of $\sim$1.8. As no spectral variability could be measured between the two revolutions, we also extracted the average ISGRI spectrum by combining all data available. Finally, we fit the ISGRI spectrum in the energy band 20 keV -- 100 keV.

\section{Results}
\label{spec_analys}

\subsection{Persistent emission}

\subsubsection{9th October}
\label{9oct}
We started our analysis with the \xmm\ data, taken during the beginning of the decaying phase (9th October). Because the first part of the ISGRI observations was taken on the same epoch and no significant spectral variability was found in these data, in order to have a broad-band spectrum we fitted simultaneously the EPIC-pn/RGS and ISGRI data. 

On the basis of the spectral properties of the previous outburst, we initially adopted an absorbed {\sc nthcomp} model \citep{zdiarski96,zycki99} in {\sc xspec} (v. 12.8.2; \citealt{arnaud96}). For the absorption, we used the {\sc tbabs} model with the abundances of \citet{wilms00}. We also added a multiplicative constant to take into account inter-calibration uncertainties between the different instruments and flux variations. We note that in the EPIC-pn spectrum there was a strong instrumental Au emission line (2.2 keV) and we fitted it with a gaussian in all the adopted models.

The {\sc tbabs$\times$nthcomp} fit did not provide acceptable results ($\chi^2/dof=969.12/764$), leaving several residuals especially in the band 1--10 keV. 
\begin{figure*}
\center
\includegraphics[width=6.5cm,angle=270]{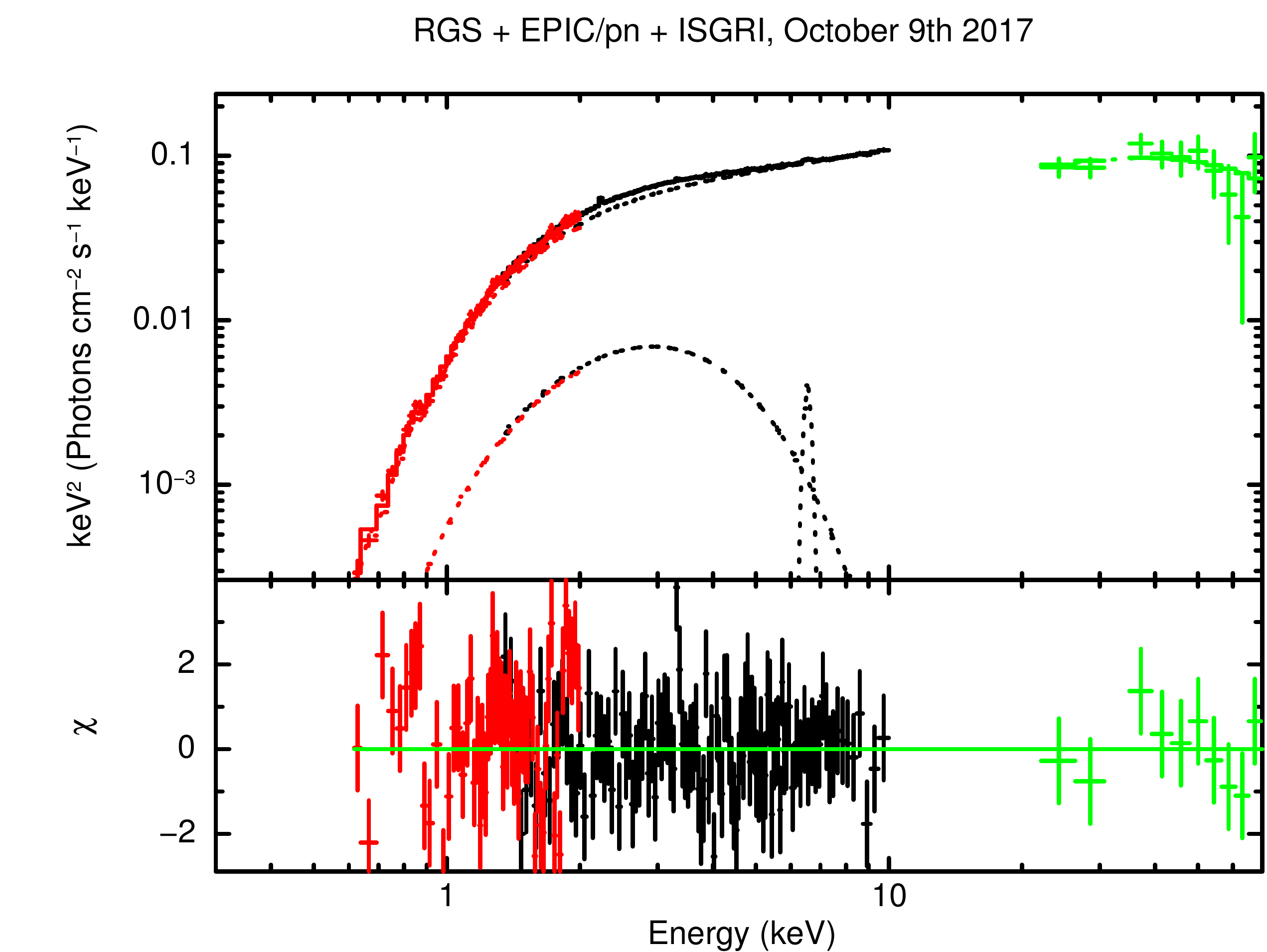}
\includegraphics[width=6.5cm,angle=270]{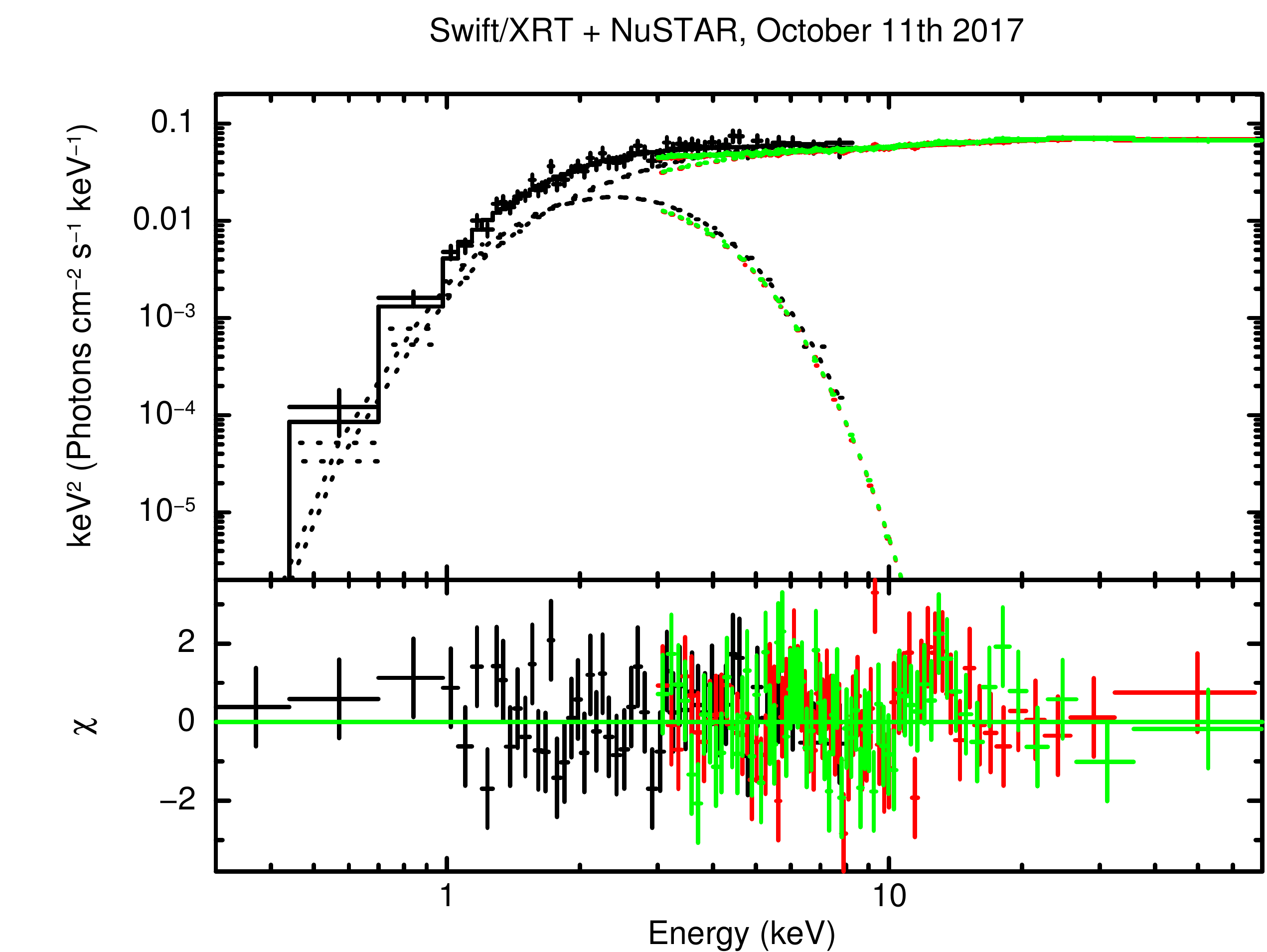}
\caption{Left: unfolded $E^2f(E)$ \xmm/EPIC-pn (\textit{black}), RGS (\textit{red}) and ISGRI ({\it green}) spectra fitted with an absorbed {\sc bbodyrad + nthcomp + gaussian} model. Right: \swift\ (\textit{black}), \textit{NuSTAR}/FMPA and FMPB (\textit{red} and \textit{green}), fitted with an absorbed {\sc bbodyrad + nthcomp} model. Data have been rebinned for display purposes only.}
\label{lc_swfit}
\end{figure*}
Therefore, we added a soft component (a {\sc diskbb} model; \citealt{mitsuda84}) that significantly improved the best-fit $\chi^2$ (899.87 for 763 dof). Its best-fit normalization, for a temperature of $0.76\pm0.04$ keV and for an inclination angle of $\sim40$\textdegree\ (as found in \citealt{pintore16}),  implies an implausible inner accretion disc radius of $\sim1.7-3.1$ km. Hence, this was more likely modelling the emission from the NS surface and we changed the {\sc diskbb} with a {\sc bbodyrad} model that, as expected, did not change significantly the best fit and gave a temperature of $0.67\pm 0.06$ keV and an emitting radius of $2.4\pm0.2$ km. 

The sp residuals suggested to add also an emission line at the energy of the Iron K-shell transitions (6.4--6.9 keV), that we modelled with a gaussian. The Iron line has an energy of 6.5 keV, a width ($\sigma$) of 0.1 keV (although the uncertainties are large) and an equivalent width (EW) of 0.015 keV, which we suggest may be due to the reflection off of hard photons from the inner regions of the accretion disc \citep[e.g.][]{fabian89}. We then tried to substitute the {\sc gaussian} model with a {\sc diskline} model \citep{fabian89}. Adopting the more recent and constrained spectral estimates reported in \citet{pintore16}, we fixed the inclination angle to 44\textdegree, the emissivity index to -2.7 and the outer disc radius at $10^5$ gravitational radii: as expected, this model does not change significantly the best fit, giving an energy line of 6.50$\pm0.15$ keV. The inner disc radius was poorly constrained (R$_{in}=43^{+31}_{-43}$ km) and no robust indication can be inferred. We tentatively adopted also self-consistent reflection models as {\sc reflionx} (\citealt{ross05}) or {\sc rfxconv} (\citealt{kolehmainen11}) but they did not improve the fit and the spectral parameters could not be well constrained.

\begin{table}
\footnotesize
\begin{center}
\caption{Best fit spectral parameters obtained with the {\sc bbodyrad + nthcomp + gaussian} model. Errors are at 90$\%$ for each parameter.}
\label{table_continuum}
\begin{tabular}{llll}
\hline
 & & 9th Oct. & 11th Oct. \\
Model & Component & {XMM+ISGRI} & {XRT+NuSTAR} \\
\\
{\sc TBabs} & n$_{H}$ ($10^{22}$ cm$^{-2}$) & $0.51^{+0.02}_{-0.02}$ &  $0.62^{+0.14}_{-0.12}$\\
\\
{\sc bbodyrad} & kT$_{bb}$ (keV) & $0.72^{+0.04}_{-0.06}$ & $0.54^{+0.04}_{-0.04}$ \\
 & norm & $6.0^{+1.1}_{-1.3}$ & $29^{+15}_{-9}$ \\
\\
{\sc nthComp} & $\Gamma$ (XMM) & $1.62^{+0.03}_{-0.04}$ & $1.79^{+0.01}_{-0.01}$ \ \\
 & kT$_{seed}$ (keV) & $0.43^{+0.03}_{-0.03}$ & = kT$_{bb}$ \\
 & kT$_e$ (keV) & $17^{+9}_{-4}$ & $23^{+3}_{-4}$ \\
 & norm ($10^{-2}$) & $1.98^{+0.07}_{-0.07}$ & $1.2^{+0.2}_{-0.2}$ \\
\\
{\sc gaussian} & Energy (keV) & $6.54^{+0.08}_{-0.07}$ & 6.54$^*$ \\
 & $\sigma$ (keV) & $0.13^{+0.13}_{-0.09}$ & 0.13$^*$ \\
 & norm (10$^{-5}$) & $3.3^{+1.7}_{-1.3}$ & $<2.1\times10^{-5}$$^{**}$ \\
 & EW (keV) & 0.015 & 0.004 \\
\hline
		 	    &$\chi^2/dof$				   &875.85/774	& 931.63/936			\\
\hline
\end{tabular}
\flushleft $^*$ fixed to this value.\\ $^{**}$ 90$\%$ upper limit on the normalization.
\end{center}
\begin{flushleft} 
\end{flushleft}
\end{table}

The final best-fit model ({\sc bbodyrad + nthcomp + gaussian}; $\chi^2/dof=875.85/774$) provided the best-fit parameters reported in Table~\ref{table_continuum} and shown in Figure~\ref{lc_swfit}-left. We note that the nH is smaller than the one reported in \citet{pintore16}. Such a discrepancy may be related to a possible variation of the local absorption during the latest outburst.
Furthermore, we mention a caveat of our best-fit model as we found a degeneracy between the spectral parameters kT$_{seed}$ and kT$_{bb}$, as acceptable fits can be also obtained with the {\sc bbodyrad} temperature smaller than the seed photon temperature. In the case the two temperatures were linked, the best-fit converges towards a mean temperature of $0.53\pm0.2$ keV without a significant change in the other parameters but the statistical significance worsens. 
 
Finally, we report that the averaged, unabsorbed 0.3--70 keV flux of \src\ was $(3.6\pm0.03)\times10^{-10}$ erg cm$^{-2}$ s$^{-1}$, corresponding to a luminosity of L$_x\sim3\times10^{36}$ erg s$^{-1}$ (for a distance of 8.5 kpc), i.e. $\sim$1.6\% of the Eddington limit.

\subsubsection{11th October}
We then analyzed the simultaneous \swift\ (Obs.ID: 00010341005) and \nustar\ spectra taken on the 11th of October. As done in the previous section, we described the broad-band continuum with a {\sc bbodyrad + nthcomp} model, obtaining a statistically significant fit ($\chi^2/dof=931.24/935$). However, as the seed photons temperature was quite unconstrained and possibly consistent with the {\sc bbodyrad} temperature, we linked them in the fit (best-fit reported in Table~\ref{table_continuum} and shown in Figure~\ref{lc_swfit}-right). 
We found a significant spectral variability in the photon index ($\Gamma\sim1.8$) and blackbody temperature (kT$_{bb}\sim0.55$ keV) with respect to the 9th of October observation. We calculated that the averaged, unabsorbed 0.3--70 keV flux of \src\ was $(2.2\pm0.1)\times10^{-10}$ erg cm$^{-2}$ s$^{-1}$, i.e. $\sim40\%$ lower than the flux measured with \xmm+ISGRI. The flux on 11th October corresponds to a luminosity of L$_x\sim1.9\times10^{36}$ erg s$^{-1}$ (for a distance of 8.5 kpc), i.e. about 1\% of the Eddington limit. 

Furthermore, we note that there is no evidence for an Iron emission line. Adding a gaussian line with energy and width fixed to those found in the previous observation, we estimated a 90$\%$ upper limit on the line normalization of $2.1\times10^{-5}$ photons cm$^{-2}$ s$^{-1}$ and on the equivalent width of 0.004 keV.

\subsubsection{5th-14th October}
Finally, we analyzed the average spectrum of each \swift\ observation during the period 5th-14th October. We fitted them, in the range 0.3--10 keV, with a {\sc tbabs$\times$nthcomp} model where we only fixed to 20 keV the electron temperature (as the fits were insensitive to this parameter). We note that there was no statistical need to add a soft component in any observation.
The single {\sc nthcomp} model provided very good results in all cases ($\chi^2/dof\sim1$), where we found that the inferred spectral parameters (nH, $\Gamma$ and kT$_{seed}$) were all consistent within 2$\sigma$ with those reported in Table~\ref{table_continuum}, implying no remarkable spectral variability.

\subsection{Type-I burst}

\src\ is a well known type-I X-ray burster and, during the 2017 outburst, we found at least two type-I bursts, one during the first \swift\ observation (at 19:32:14 UTC on October 5th) and one during the \xmm\ observation (at 21:05:11 UTC on October 9th). In addition, MAXI/GSC detected a bright X-ray burst at 06:44 UTC on October 6th, which lasted for 10 s and consistent with the position of \src\ \citep{harita17}.

Because of the higher data quality, we focus on the burst observed in \xmm. Modelling the burst lightcurve with an exponential function plus a constant, we found that the e-folding decay time is $25\pm0.5$ s, similar to those inferred from previous \src\ type-I bursts.

\begin{figure}
\center
\includegraphics[width=11.9cm]{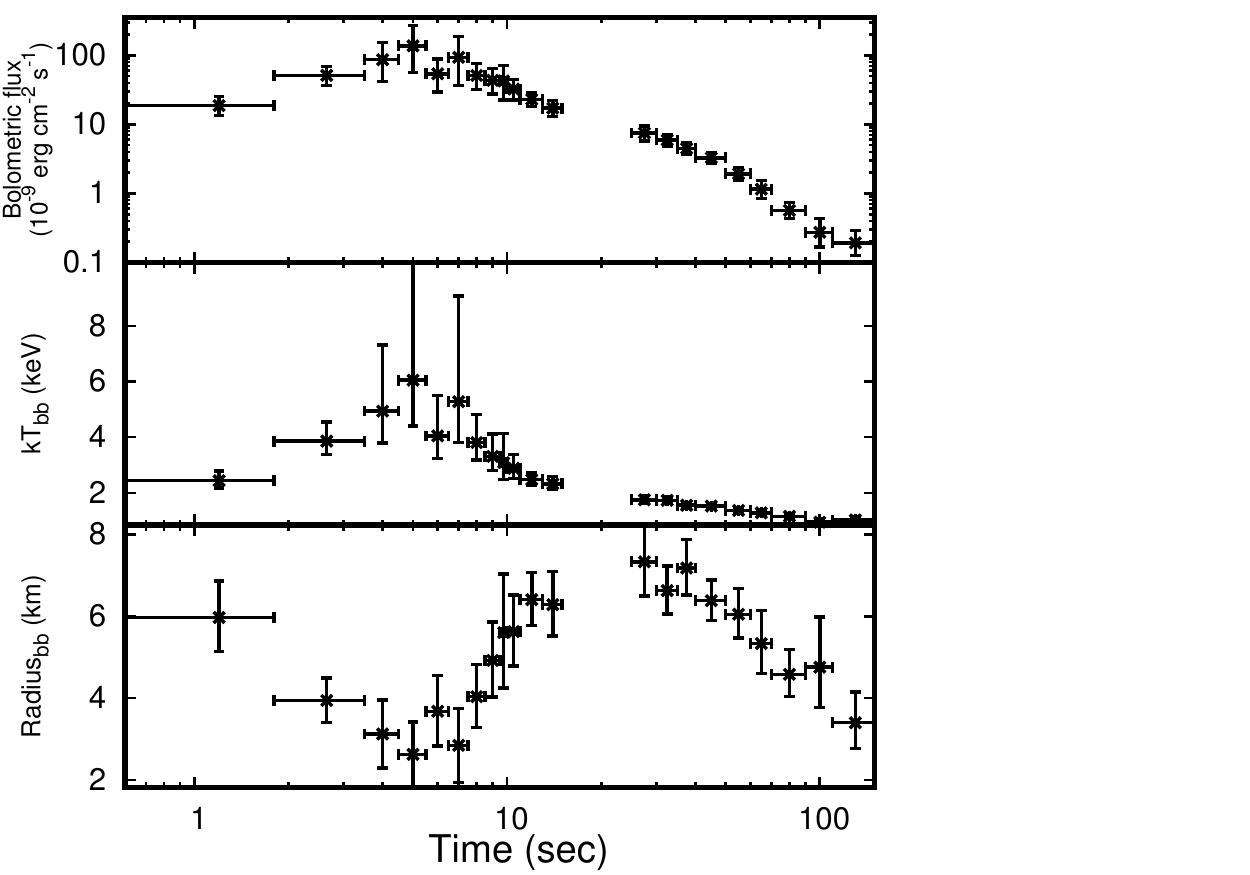}
\vspace{-0.5cm}
\caption{Bolometric flux, blackbody temperature and radius profiles of the type-I X-ray burst occurred during the \xmm\ observation. The gap at $\sim12$s--25s is due to a temporary outage of the EPIC-pn instrument.}
\label{burst_par}
\end{figure}

We then carried out a time-resolved spectroscopy, selecting a number of time-intervals from the beginning of the burst to $\sim130$s after it. For each interval, we extracted a spectrum (corrected for pile-up by excluding the central column RAWX=37 from the data) and we fitted it with an absorbed {\sc bbodyrad} model (fixing n$\text{H}$ to $0.51\times10^{22}$ cm$^{-2}$) in the range 1.0--10 keV, where we used as background the persistent emission pre-burst. The bolometric flux (estimated according to equation 3 in \citealt{galloway08}), the best-fit blackbody temperature, and the corresponding emitting radius are shown in Figure~\ref{burst_par}. The profile shows that the flux increased, peaked at a constant value for $\sim$10s, and then it decreased to the persistent emission level. A similar trend is observed for the temperature as well, which reached up to a maximum of 5 keV (although with large uncertainties). Instead, the radius of the emitting regions has a more complex trend that resembles a photospheric radius expansion (PRE) event, which is not unusual for this source \citep{galloway08,guver13}. The mean bolometric flux registered at the peak (which lasted $\sim$10 s) with \xmm\ was $(50\pm4) \times 10^{-9}$ erg cm$^{-2}$ s$^{-1}$. 

\section{Discussion}

In this work we present the 2017 outburst of \src\ which lasted for $\sim13$ days and with an exponential decay with e-folding decay time of $\sim$4 days. According to the {\it MAXI} lightcurve, \src\ reached the peak of the outburst between October, 4 and 5. Since the first \swift\ observation was taken on October, 5, we estimate that \src\ possibly reached at least an absorbed 0.3--70 keV peak flux of $(7.0\pm0.7)\times10^{-10}$ erg cm$^{-2}$ s$^{-1}$, i.e. a factor of 2 higher than the flux measured during the latest \xmm\ observation. This flux corresponds to a luminosity of L$_x\sim6.0\times10^{36}$ erg s$^{-1}$ (for a distance of 8.5 kpc), hence about 4$\%$ of the Eddington luminosity. 

Taking into account the marginal spectral variability during the decay, we found that the spectral properties of the 2017 outburst were characterized by a strong comptonized emission with electron and seed photon temperatures of $\sim17-23$ keV and $0.4$ keV, respectively, and photon index of 1.6--1.8. Such values correspond to an optical depth of $\tau\sim3-5$ (see equation A1 in \citealt{zdiarski96}). We estimated that, for a Compton parameter $y\sim3$, a bolometric flux of $3.4\times10^{-10}$ erg s$^{-1}$ and a distance of 8.5 kpc, the radius of the region responsible for the seed photons is about 12 km (adopting the approach showed in the Discussion section of \citealt{intZand99}).  
It is therefore likely that the seed photons are produced in a region close to the NS, as the boundary layer or the NS surface, although we cannot rule-out contamination also from the accretion disc. 

Furthermore, we found evidence of a significantly weaker blackbody component (although we remind here about the degeneracy between blackbody and seed photon temperature in the spectral fits discussed in Section~\ref{9oct}) with a temperature of $\sim0.7$ keV and emitting radius of a few km, which origin is likely from the NS surface. This component carried $\sim$$5\%$ and $\sim$$11\%$ of the total source luminosity on October, 9 and October, 11, respectively.

On October, 9, it was also observed the presence of a relatively broad Iron line at $6.5$ keV. We exclude that the broadening is either compatible with Compton down-scattering in a wind shell ejected at moderate relativistic velocities from the disc \citep{laurent07} or Compton processes in an accretion disc corona above the accretion disc \citep{white82a,kallman89, vrtilek93}. Indeed, should these processes responsible for the broadening, the corresponding electron temperature can be obtained from the relation $\Delta \epsilon/\epsilon= (4kT_e-\epsilon)/m_ec^2$ (where $\Delta \epsilon$ is the line broadening and $\epsilon$ is its energy). However, we estimated an electron temperature of $\sim4$ keV which is a factor of 4--5 lower than that found from our spectral analysis. Even if we consider the 90$\%$ upper uncertainties on the broadening and energy line (i.e. 0.26 keV and 6.62 keV), we inferred an upper limit on the electron temperature of $\sim7$ keV, still well below our value. For this reason, we associate the line broadening to reflection processes from the surface of the accretion disc \citep[see e.g.][]{fabian89}. Using the {\sc diskline} model, we could not constrain the dimension of the inner disc radius, and thus no definitive conclusion can be inferred on this parameter. Similar results are also obtained with self-consistent reflection models.

We remark that this is the sixth registered outburst of \src, but its maximum peak luminosity is the weakest ever reported for this source to date (although {\it Swift} may underestimate the broadband source flux). In fact, all the other five events were significantly brighter, with the 1998 one being the shortest in duration and the lowest in peak luminosity \citep{altamirano08}. However the latter is still at least a factor of 3 higher than the peak luminosity of the 2017 outburst. 
For the 1998 outburst, using {\it BeppoSAX} data, \citet{intZand99} found that the 0.1--100 keV spectrum of \src\ was consistent with a single thermal Comptonization model with electron temperature of 15 keV, optical depth in the range 3--6 and seed photon temperature of 0.6 keV (coming from a region of radius of $\sim13$ km). This indicates that the source was in a {\it hard} state. Based on our analysis, we suggest that \src\ was again in a {\it hard} state during the 2017 outburst (and possibly all along the decay as the \swift\ monitoring did not show any clear evidence of spectral variability) making the two outbursts very similar. Pulsations have been mainly observed during the hard states of \src\ and we confirm a pulse detection, that we will report in greater detail elsewhere (Sanna et al. in prep.).

The last outburst was hence clearly different from the 2015 one, when the source reached $\sim25\%$ of the Eddington luminosity, and its spectral properties were consistent with a marked {\it soft} state \citep{pintore16}. In particular, it was characterized by the combination of four spectral continuum components, representing the disc emission, the NS surface, a comptonized region (possibly the boundary layer) and an additional non-thermal emission (described with a hard powerlaw), carrying $\sim20\%, 25\%,51\%$ and $3\%$ of the total luminosity, respectively. In addition, the NS surface emission in 2015 had a temperature of $\sim$$1.1$ keV, about a factor of 2 higher than the temperature estimated in this work. In 2017, we also did not find any evidence of a hard powerlaw tail at high energy, confirming the {\it hard} state of the source, as such a component is generally seen only during {\it soft} states of LMXBs \citep[e.g][]{disalvo00, disalvo01, damico01, iaria04, disalvo06, paizis06, dai07, tarana07, piraino07}. 

At least three type-I X-ray bursts were observed from \src\ with MAXI, \swift\ and \xmm\ in its 2017 outburst unlike the 1998 outburst where none of them were found. According to the linear relation between pre-burst count rate and elapsed time reported in \citet{pintore16} for the 2015 outburst, we predict that the expected recurrence time of the type-I bursts during the 2017 outburst would have been at least one every $\sim8-9$ ks, much shorter than the observed recurrence time. 
During the burst, the measured NS surface temperature increases up to $\sim5$ keV (although the uncertainties are quite large). On the other hand, the radius of the emitting area presents a more complex trend: in fact, it is $\sim6$ km at the beginning of the burst and it decreases below $\sim4$ km during the burst rising phase; at the maximum flux, the radius increased again to $\sim6$ km, remaining constant for at least 60 seconds. This behaviour is similar to the {\it radius expansion} process, i.e. the response of the NS outermost layers to a super-Eddington burst flux \citep[see e.g.][]{kuulkers03}. In particular, the NS photosphere increases and expands and then contracts back to the NS surface. It was found that during the expansion/contraction phase the flux remains nearly constant to the Eddington limit, converting any excess into kinetic energy of the outflow \citep[e.g.][]{kato83,ebisuzaki83,paczynski86}. We note that radius expansions are not unusual for \src\ \citep[e.g.][]{guver12_2} although in the 2015 outburst they were not observed in any of the detected type-I X-ray bursts. 
Assuming that the mean flux at the peak of the burst reported here ($5.0\times10^{-8}$ erg cm$^{-2}$ s$^{-1}$) reached the Eddington limit and using the Eddington luminosity empirically obtained by \citet{kuulkers03} ($3.8\times10^{38}$ erg s$^{-1}$), we estimated a source distance of $8.0\pm0.4$ kpc and $5.2\pm0.4$ kpc (error at 1$\sigma$) for a pure helium ($X=0$) or a solar composition burst ($X=0.7$), respectively, which are well consistent with the previous distance estimates \citep{galloway08}.

\section*{Acknowledgements} 

We thank both Dr. F. Harrison and Dr. N. Schartel, who made these ToO observations possible using their Director Discretionary Time, and the {\it XMM-Newton} and the {\it NuSTAR} team who performed and supported these observations.
We used observations obtained with XMM-Newton, an ESA science mission with instruments and contributions directly funded by ESA Member States and NASA.
We acknowledge financial contribution from the agreement ASI-INAF I/037/12/0.
We acknowledges support from the HERMES Project, financed by the Italian Space Agency (ASI) Agreement n. 2016/13 U.O, as well as fruitful discussion with the international team on ``The disk-magnetosphere interaction around
transitional millisecond pulsars at the International Space Science Institute, Bern''.

\addcontentsline{toc}{section}{Bibliography}
\bibliographystyle{mn2e}
\bibliography{biblio}

\end{document}